\documentclass[conference]{IEEEtran}
\pdfoutput=1
\IEEEoverridecommandlockouts

\usepackage{tikz}
\usepackage{textcomp}
\usepackage{lipsum}

\newcommand\copyrighttext{%
  \footnotesize \textcopyright 2021 IEEE. Personal use of this material is permitted.
  Permission from IEEE must be obtained for all other uses, in any current or future
  media, including reprinting/republishing this material for advertising or promotional
  purposes, creating new collective works, for resale or redistribution to servers or
  lists, or reuse of any copyrighted component of this work in other works.
  DOI: \href{<http://tex.stackexchange.com>}{<DOI No.>}}
\newcommand\copyrightnotice{%
\begin{tikzpicture}[remember picture,overlay]
\node[anchor=south,yshift=10pt] at (current page.south) {\fbox{\parbox{\dimexpr\textwidth-\fboxsep-\fboxrule\relax}{\copyrighttext}}};
\end{tikzpicture}%
}

\usepackage{cite}
\usepackage{float}
\usepackage{subcaption}
\usepackage{dblfloatfix}
\usepackage{amsmath,amssymb,amsfonts}
\usepackage{algpseudocode}
\usepackage{algorithm}
\usepackage{array,multirow,graphicx}
\usepackage{textcomp}
\usepackage{xcolor}
\def\BibTeX{{\rm B\kern-.05em{\sc i\kern-.025em b}\kern-.08em
    T\kern-.1667em\lower.7ex\hbox{E}\kern-.125emX}}
\usepackage{booktabs}
\usepackage{enumitem}
\usepackage{threeparttable}
\usepackage[multiple]{footmisc}
\usepackage{balance}
\usepackage{xcolor,colortbl}
\usepackage{todonotes}
\usepackage{tabularx}
\usepackage{hyperref}
\hypersetup{colorlinks,allcolors=black}

\begin{document}

\title{Dependable IoT Data Stream Processing for Monitoring and Control of Urban Infrastructures}

\author{
\IEEEauthorblockN{Morgan K. Geldenhuys, Jonathan Will, Benjamin J. J. Pfister,\\Martin Haug, Alexander Scharmann, and Lauritz Thamsen}
\IEEEauthorblockA{Technische Universit{\"a}t Berlin, Germany, \{firstname.lastname\}@tu-berlin.de}
}

\maketitle
\copyrightnotice

\makeatletter

\begin{abstract}
The Internet of Things describes a network of physical devices interacting and producing vast streams of sensor data. 
At present there are a number of general challenges which exist while developing solutions for use cases involving the monitoring and control of urban infrastructures. 
These include the need for a dependable method for extracting value from these high volume streams of time sensitive data which is adaptive to changing workloads.
Low-latency access to the current state for live monitoring is a necessity as well as the ability to perform queries on historical data.
At the same time, many design choices need to be made and the number of possible technology options available further adds to the complexity.

In this paper we present a dependable IoT data processing platform for the monitoring and control of urban infrastructures. 
We define requirements in terms of dependability and then select a number of mature open-source technologies to match these requirements.
We examine the disparate parts necessary for delivering a holistic overall architecture and describe the dataflows between each of these components. 
We likewise present generalizable methods for the enrichment and analysis of sensor data applicable across various application areas. 
We demonstrate the usefulness of this approach by providing an exemplary prototype platform executing on top of Kubernetes and evaluate the effectiveness of jobs processing sensor data in this environment.

\end{abstract}


\section{Introduction}

The Internet of Things (IoT) has emerged as an important technological paradigm whereby large numbers of ubiquitous devices are networked to enable the real-time monitoring and control of physical infrastructures across a wide range of application areas. 
IoT owes its existence to the convergence of a number of key enabling technologies such as ubiquitous computing, commodity sensors and micro-controllers, machine-learning, and real-time analytics~\cite{Gubbi2013InternetOT}. 
Important areas of application include the management of critical infrastructures such as human health-care, transportation systems, electrical generation, natural disaster prediction, and telecommunications, to name but a few~\cite{GJF+16,JGM+14,CLC+15,LPS16}.
With the number of sensor devices only projected to increase, i.e. from 10 billion in 2021 to more than 25 billion devices in 2030\footnote{https://www.statista.com/statistics/1183457/iot-connected-devices-worldwide/, Accessed: Jul 2021}, it follows then that data volumes will likewise reach ever great heights in the years to come.
Therefore, in order to meet processing demands a dependable IoT processing platform is necessary to ensure results are available when time-sensitive decisions need to be made as processing loads invariably increase over time.


However, building such a processing platform for geo-distributed urban infrastructures is not an easy task.
There are a number of requirements, design choices, and technological options which need to be considered.
Not only should the platform provide fast access to results describing the current state of the infrastructure to enable live monitoring, but also allow for the storage and analysis of historical data to identify trends over longer periods of time.
Results should be presented in a format which is understandable to both the user and analysis jobs which at the same time ensures low-latency transmission of sensor data in an environment where network resources are limited.
Likewise, there are a multitude of technology options from which to choose from for the fulfilment of individual use cases. 
This logically begs the question: which set of systems should be chosen so that when they are composed together they will produce a dependable solution?

Selecting the right technologies is not the only problem, such a platform consisting of complex inter-dependencies will likewise be difficult for operators of the platform to deploy and maintain.
A number of solutions have been proposed in this area, however they tend to be specific to application domains not generalizable to urban infrastructures~\cite{zeuch2020nebulastream,ye_hydrologic_2020,difallah_scalable_2013,kolozali_knowledge-based_2014} or do not provide data about the scalability and/or reliability of their platform~\cite{malek_use_2017,huru_bigclue_2018}.
Likewise, the number of commercially available end-to-end IoT solutions for the processing of large data streams is growing\footnote{https://cloud.google.com/solutions/iot, Accessed: Jul 2021}\footnote{https://aws.amazon.com/iot/, Accessed: Jul 2021}\footnote{https://www.ibm.com/cloud/internet-of-things, Accessed: Jul 2021}, however, these tend to be proprietary with little regard for open standards as vendors compete in the already lucrative IoT market.
Additionally, such solutions are not feasible when the requirement for on-premise operations exist due to privacy and/or data protection issues.


In this paper we present a dependable data processing platform for IoT sensor streams. 
This open source platform is targeted towards the monitoring and control of urban infrastructures.
We investigate and define the requirements for creating such a platform in the context of dependability with specific focus on availability, reliability, maintainability, and security.
We propose an architecture consisting of a number of interconnected sub-systems which performs all necessary functions expected of a IoT data processing platform as well as ensuring that it is fast to deploy and easy to maintain.
Likewise, we present two stream processing jobs essential to performing enrichment of sensor values as well as analysis as a generalizable use case for geo-distributed urban infrastructures.
We perform experiments to evaluate the scalability and reliability of this setup and present the results.
Our evaluation reveals our analysis platform architectures exhibits near-linear scaling capabilities for realistic example big data workloads.


The remainder of the paper is structured as follows: Section II presents an investigation of the requirements in the context of dependability; Section III presents the anatomy of the IoT data processing platform; Section IV details two stream processing jobs which are generalizable across most IoT monitoring and control use cases; Section V describes our evaluation through experiments; Section VI the related work; and Section VII summarizes our findings.

\section{Platform Requirements}

In this section we will discuss the requirements of a dependable IoT data processing platform.
The function of such a platform is to consume the incoming IoT sensor data in order to extract valuable results by cleaning, filtering, enriching, aggregating, and analysing the time-sensitive data streams.
Importantly, stream processing is concerned with providing high rates of throughput, low latency processing, and fault tolerant execution.
Therefore, it is commonplace for constraints to exist which dictate the minimum performance requirements.
For a platform of this type to be dependable, it should fulfil the following requirements.
\begin{itemize}
  \item The platform should deliver the service that is required considering the presence of any performance constraints and adapt to any workload changes. For this to be true, components of the system need to be scalable.
  \item The platform should be able to continue operating in the presence of partial failures. Therefore, it is important that components are fault tolerant and able to recover automatically when failures occur. 
  \item Distributed systems are notoriously complex and composing multiple of these systems together only exacerbates the problem. Therefore, the platform should abstract away this complexity from users making it relatively straightforward to use and maintain. 
  \item The platform should be secured against malicious threats. Therefore, all externally facing interfaces should implement strict security measures and the number of these should be minimized to reduce the attack surface.
\end{itemize}

\begin{figure*}
    \centering
    \includegraphics[width=0.8\textwidth]{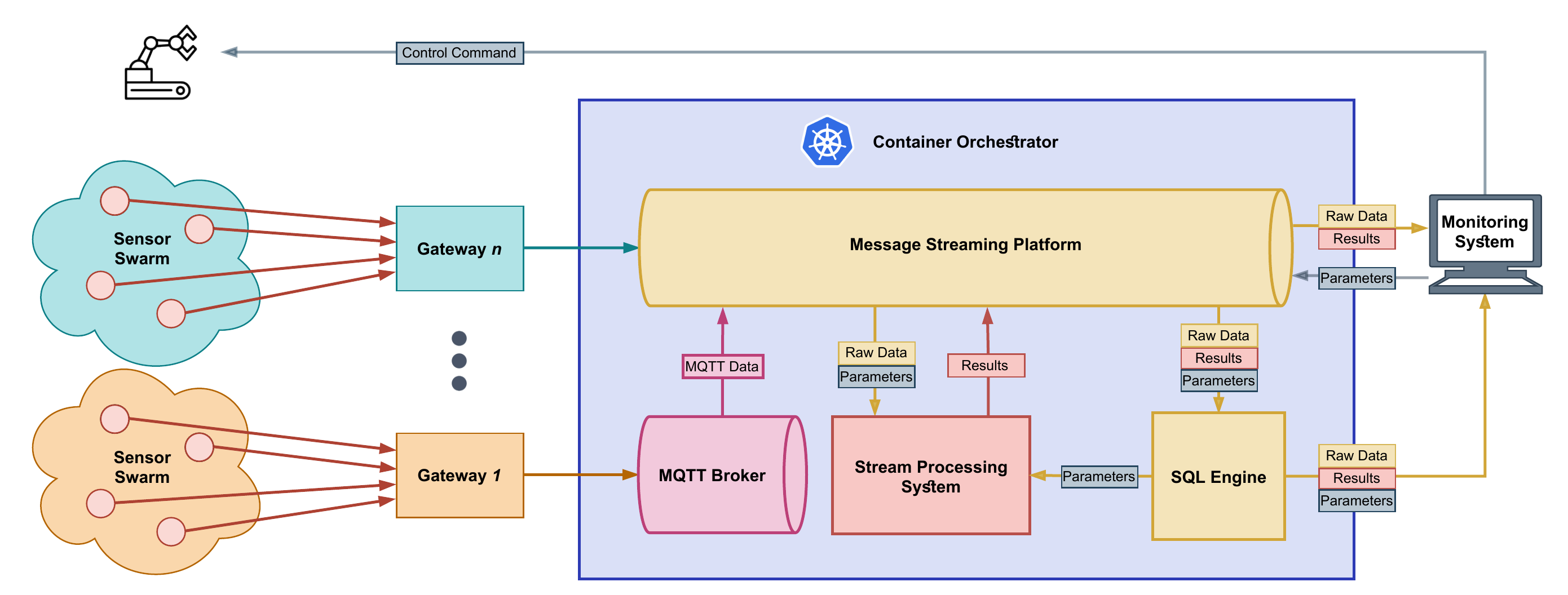}
    \caption{Dependable IoT Data Processing Platform Architecture.}
    \label{platform}
\end{figure*}

\section{Platform Anatomy}

Following on from the requirements presented in the previous section, here we detail our proposal for a dependable IoT data processing platform. 
We compose the platform from a number of key sub-systems which are interconnected and pass sensor streaming data between them. 
A graphical representation of this platform can be seen in Fig. \ref{platform}. 
Data is generated from sensors placed in the field and flows through the platform where results are ultimately consumed by the monitoring system. 
The platform allows for both the consumption of live streaming data which is vital when considering the current state monitoring of a physical infrastructure, as well as, the capability to perform precision queries of historical data.
Importantly, all components should provide the ability to secure outward facing communication channels to ensure the confidentiality and integrity of data.
All sub-system recommendations mentioned in this section provide this functionality through encryption.
Next we will describe the individual components of this platform and how they are intended to interact with one another.

\subsection{Sensors and Gateways}

Data is generated from sensors which are in turn connected to devices. Multiple sensors can be connected to the same device and each measures a specific physical quality, i.e. light intensity, distance, temperature, acceleration, etc. Oftentimes the devices are low powered microcontrollers with limited processing capacity. These devices will transmit the data to a gateway which acts as a bridge between the data generators and data processing environment. As sensor devices are generally remotely placed, they often use a variety of wireless protocols such as LoRa\footnote{https://lora-alliance.org/, Accessed: Jul 2021} or Zigbee\footnote{https://zigbeealliance.org/about/, Accessed: Jul 2021} to transmit data and therefore require some intermediate infrastructure to pass data on to a TCP/IP based network. Examples of such infrastructures include ChirpStack\footnote{https://www.chirpstack.io/, Accessed: Jul 2021} and The Things Network\footnote{https://www.thethingsnetwork.org/, Accessed: Jul 2021}.
Regarding data formatting, it is important to reduce the amount of data which is transferred over these low bandwidth network.
Therefore, the sensor network should transmit only the sensor ID, timestamp, and value while relying on the analytics platform to perform any enrichment and/or calibration of the data.
A specification such as SenML should be used across the whole infrastructure for the encoding of sensor measurements and device parameters used in enrichment.
For use cases where control functionality is necessary, devices are able to accept commands from some human-machine interface.

\subsection{Distributed Message Streaming Platform}

At the heart of the IoT data stream processing platform is a scalable message broker. This distributed middleware component allows for data to be published to and consumed from messaging queues. 
To ensure a high level of performance is maintained at any scale, all data should be routed through this platform as it decouples components from each other where slow sinks are a regular limiting factor. 
Therefore, instead of a pipeline of systems, a more star-like architecture is adopted with the message broker at the center. 
To ensure that this component does not itself become a performance bottleneck, it must exhibit two key characteristics: (1) the ability to replicate copies of the message queues across multiple brokers thus ensuring failure tolerance, and (2) the ability to partition individual message queues so that multiple sinks and sources can consume/publish events in parallel. 
An example of such a system would be Apache Kafka~\cite{Sax2019ApacheK}.
Importantly, a system like Kafka can be configured to store individual message queues indefinitely and as such essentially becomes a permanent datastore.
This setup is inherently scalable as well as fault tolerant.
However, traditional database systems have the advantage of being able to perform intricate queries over the data which is normally not possible with these types of systems.
Kafka, however, when combined with a SQL engine like ksqlDB\footnote{https://ksqldb.io/, Accessed: Jul 2021} is able to provide this functionality.
KsqlDB is a database which operates on top of Kafka which is purpose-built for stream processing use cases.

\subsection{MQTT Broker}

This optional component can be deployed if support for the Message Queuing Telemetry Transport (MQTT) protocol is required. 
As systems like Apache Kafka do not support MQTT, we can instantiate an intermediary broker as a target for MQTT data producers. 
RabbitMQ\footnote{https://www.rabbitmq.com/, Accessed: Jul 2021} is an example of such a system. Messages are produced from the MQTT gateway into this broker after which they are forwarded to the distributed message streaming platform where they are be processed as per normal.

\subsection{Distributed Stream Processing System}

DSP systems are becoming an increasingly essential part of any data processing environment. 
It is here where messages must traverse a graph of streaming operators to allow for the extraction of results which are at their most valuable at the time of data arrival. 
Such systems allow users to define streaming jobs and run them upon a cluster of commodity nodes which is able to scale out to process any potential workload. 
It is here where user defined jobs are executed.
The raw sensor data is consumed from the message streaming platform after which it is enriched if necessary and any analysis performed. 
Results are outputted directly to the message streaming platform in order to bypass any slow sinks. 
Example systems include Apache Spark Streaming~\cite{ZMC+10} and Flink~\cite{CKE+15}.
The DSP system on startup will load any existing parameters from the SQL engine and store them in an in-memory data structure to be used during data enrichment and/or analysis processes.
This mechanism will be further discussed in the subsequent section.
It also provides a mechanism for users to update parameters during runtime execution.
The \textit{monitoring system} publishes the new parameters to the message streaming platform which in turn is consumed by the DSP and the local in-memory parameter cache updated. 
For the redundant storage of models and checkpoints, an external service such as Ceph~\cite{Weil2006CephAS} or HDFS~\cite{Shvachko2010TheHD} can be used.

\subsection{Monitoring System}

Users require a mechanism for interacting with the IoT platform where they can observe the current status of the system as well as the results of any analysis which may be performed.
It is here where Geographic Information Systems (GIS), for example, are useful in the visualization of spacial and geographic data~\cite{Burrough1998PrinciplesOG}.
By integrating both message streaming and SQL clients, the monitoring system is able to consume the enriched IoT data to have an overview of the current state and perform intricate queries on the data over longer time periods.
From here users have access to information necessary to make informed decisions about the infrastructures they are tasked with managing.
Additionally, certain decision making can be automated to allow for the existence of self-optimizing processes.
Although not reflected in Fig. \ref{platform}, monitoring of the compute resources is likewise an important function.
For this, a good solution is the deployment of a Prometheus\footnote{https://prometheus.io/, Accessed: Jul 2021} time series database as part of the platform which periodically scraps metrics from the underlying cluster nodes as well as sub-systems.
From here, a data visualization frontend like Grafana\footnote{https://grafana.com/, Accessed: Jul 2021} can be used to analyze performance.

\subsection{Resource Management System}

Resource management systems enable the automatic deployment, scaling, and management of containerized applications. 
When a system like Kubernetes~\cite{VPK+15} is combined with a package manager such as Helm\footnote{https://helm.sh/, Accessed Jun 2021} we are able to use infrastructure-as-code processes to efficiently define, install, and upgrade a complex graph of interdependent sub-systems.
These definitions can then be shared and updated with minimal effort without having to setup the individual systems manually.
It thus reduces the need for expert knowledge on each individual sub-system and greatly reduces the lead time to deployment.
Traditionally when deploying systems with a master-worker architecture on bare-metal, redundant nodes are required so that in the event of failure they can take over. 
However, the number of backups is finite and therefore can only tolerate a certain number of failures before the job ultimately stops processing.
With Kubernetes and Flink native\footnote{https://ci.apache.org/projects/flink/flink-docs-release-1.13/docs/ deployment/resource-providers/native\_kubernetes/, Accessed: Jul 2021}, for instance, where the stream processing system can communicate directly with the resource manager, it can now withstand an infinite number of failures by requesting new containers on demand.  
For the purposes of this paper we have created a github repository containing a helm chart for the infrastructure we describe in this section\footnote{https://github.com/morgel/tdis21\_dsp\_platform, Accessed: Jul 2021}.
\section{General IoT Jobs} 

In this section we define two generalizable DSP jobs for the monitoring and control of urban infrastructures. These jobs rely on the architecture as described in the previous section to be in place. The first is the enrichment of IoT data which is necessary when using low bandwidth wireless infrastructures, thus ensuring low-latency processing.
The second type of job uses complex event processing to allow, for instance, analyzing sensor measurements across wide geographical areas. 

\subsection{IoT Data Enrichment Job}

Server side data enrichment is a key function of any stream processing platform when trying to optimize network resource utilization.
This subsection describes how data transmitted using minimized formats can be enriched with information needed by the user in order to obtain an overview of the current state of the infrastructure as well as data analytics.

Inbound sensor data enters the stream processing system to be available for live monitoring and analytics. 
However, IoT sensors may transmit a variety of formats and data fields.
For instance, a system might have a mix of IoT devices with digital and analog temperature sensors that can either directly emit a temperature or merely the sensor's voltage reading.
The incoming data may also not be sufficient for monitoring or analysis, e.g. containing only an ID instead of the geographical latitude / longitude corresponding to this ID.
Now, instead of having to deal with multiple formats in the analytics application and having to fetch missing data from a database, we perform a \emph{data enrichment} job first.
In this job, the stream processing system digests all inbound sensor data to transform it into a single, standardized format and annotates it with all information that may be needed for further consumption. 
These enriched packets are emitted back to the message streaming platform in a separate topic that analytics tasks can subscribe to.

Bandwidth conservation is important for low-power IoT devices to save energy.
One way to achieve it, is to omit redundant data that can later be re-added to the datapoint.
Additionally one can use a minimal data notation for transmission from the IoT device to the stream processing system, as defined by e.g. the \emph{Sensor Measurement Lists}\footnote{https://www.iana.org/assignments/senml/senml.xhtml, \\Accessed: Jul 2021} (SenML).
The following exemplifies sensor data in SenML notation.
\begin{verbatim}
{"n" : "70B3D5499073C7C0-temp",
 "t" : 1625222417,
 "v" : 4.2}
\end{verbatim}
The first step to enriching this data is to look up the meaning of each key in a lookup table.
We can deduct that in this example, we see a temperature measurement from an analog sensor with the name ``70B3D5499073C7C0-temp'', taken at Unix time 1625222417 and valuing at 4.2 Volts.
After calibration, the enrichment job can assign a temperature to the voltage measurement, e.g. 23°C for 4.2 volts.
Next, we can look up the geolocation of ``70B3D5499073C7C0-temp'' from a database and append it to the data point before ingesting it back into the message streaming system.
The fully enriched data point might now look as follows:
\begin{verbatim}
{"latitude" : 52.51017262863814,
 "longitude": 13.322876673244508,
 "time" : 1625222417,
 "temperature" : 23}
\end{verbatim}
One example application in the real world that makes use of IoT data enrichment is predictive pumping of water in urban sewage systems.
Decisions here are based on continuous streams of data from geo-distributed sensors measuring suction chamber levels of water pumps on one hand and expected precipitation in an area of a city on the other hand.
In this example, sending the data in a reduced format mitigates the load on the networking of the IoT devices attached to the sensors, while the enriched data retains full data interpretability by the analytics jobs that influence the pump controls.

To summarize, the enrichment job enables platform designers to appropriately separate data ingestion and formatting concerns from the actual analysis~\cite{lorenz2020scalable}.

\subsection{IoT Analytics Job}
\label{sec:prox-cep}

Once all data arrived in the central message streaming platform and was transformed to a standardized format, it has to be analyzed to extract insight.
For this, we use \emph{Complex Event Processing} (CEP).
With this technique, the user specifies \emph{event patterns} for data streams. Patterns allow to reason about a sequence of data packets in the stream and their relations, e.g. whether a value increases or exceeds a threshold multiple times within some time frame.
A typical CEP pipeline would proceed as follows:
\begin{enumerate}
    \item Event patterns are applied to an inbound enriched data stream, creating a new stream of possible events.
    \item For each match an event is emitted to the output stream.
    \item These event streams can then be transformed to alert events to be written back into the message streaming platform for further usage.
\end{enumerate}

Buchmann et al.~\cite{buchmann2009cep} state the two advantages that make CEP a good fit for a stream processing:
Firstly, it separates analysis from the inbound data source and the outbound alert sink, thus offe nbring flexibility with regards to changes of the data producers in the urban infrastructure.
Secondly, expressive event patterns make them especially suited to analyzing streams of data as opposed to looking at single data packets in an isolated fashion.
CEP makes emitting alerts based on repeated deviation from a given value easy.
However, many real-life systems do not have a single ``normal'' value to look out for.
We examine this, using the example of sensors in street drainage with the task of reporting clogged inlets. 
The sensors observe water levels inside the inlet.
The reference value depends on weather and other events.
Assuming that most inlets are okay, we can obtain an estimate for the normal sensor value by averaging over all sensors in an area. 
The allowable sensor values can then only deviate by some fixed amount from that average, otherwise, an alarm is emitted.



\begin{figure}
    \centering
    \begin{subfigure}[t]{0.28\columnwidth}
        \includegraphics[width=\textwidth]{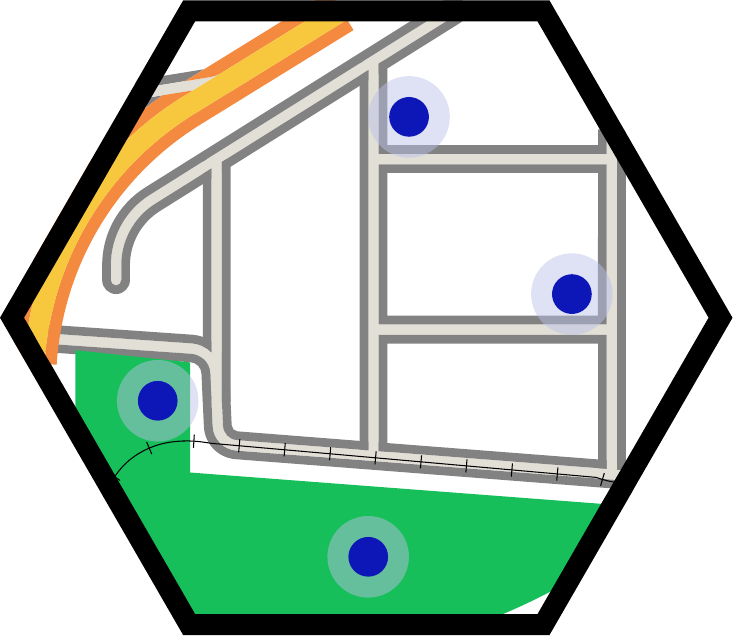}
        \caption{All sensors reporting within rolling average}
        \label{fig:cell-1}
    \end{subfigure}
    \hfill 
    \begin{subfigure}[t]{0.28\columnwidth}
        \includegraphics[width=\textwidth]{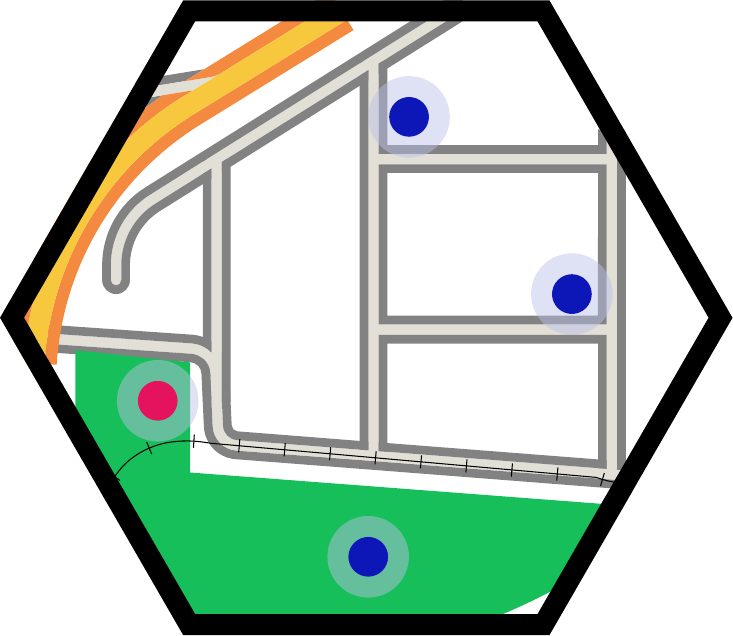}
        \caption{Anomaly type 1: One sensor anomalous}
        \label{fig:cell-2}
    \end{subfigure}
    \hfill 
    \begin{subfigure}[t]{0.28\columnwidth}
        \includegraphics[width=\textwidth]{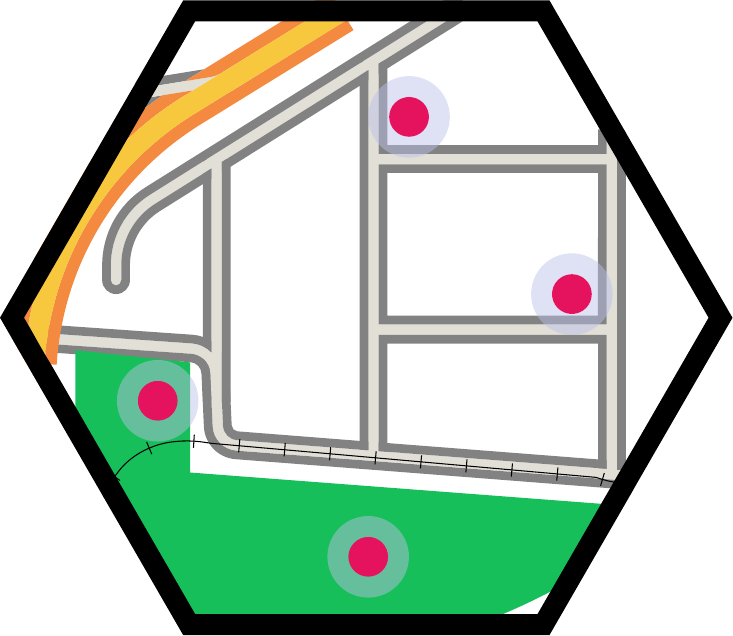}
        \caption{Anomaly type 2: All sensors suddently anomalous}
        \label{fig:cell-3}
    \end{subfigure}
    \caption{Proximity-based CEP uses a band around the rolling averages of sensors (dots) in a geographic cell. Red coloring indicates values outside of that band.}\label{fig:cep-cells}
\end{figure}

We minimize the impacts of a faulty inlet on the reference value by calculating it as a rolling average. This offers two advantages: First, when a single anomaly starts, the system does not suddenly become less sensitive because that anomaly was already fully entered into the average. Second, many systems might consider sudden changes of a value across a set of neighbored sensors of interest as well. The rolling average will not fully update to the new, anomalous average and thus, all observed sensors are now outside of the range of values around the rolling average. Fig.~\ref{fig:cep-cells} illustrates the concept.
We use Uber's H3 library\footnote{https://h3geo.org/, Accessed: Jul 2021} to partition our sensors into hexagonal cells by location. We emit an alert if a sensor reports a water level outside the range around the rolling average for its cell twice or more within a short time frame.

Another use case example for complex event processing is predictive maintenance in railway systems, in particular power transmission through pantograph-catenary systems.
Complex models that take into account data from various sensors on the pantograph and cameras on the roof are used to assess the state of the system~\cite{bruni2015results, finner2015program}.
Here, CEP enables live edge processing of data from the sensors in combination with recognized patterns from the video feed to detect anomalies\footnote{\href{https://pantohealth.com/pantosys/}{https://pantohealth.com/pantosys/}, Accessed: Jul 2021}.


\section{Evaluation}

In this section we evaluate the proposed dependable IoT data processing platform as well as both enrichment and analytical jobs through experiments designed to test scalability and reliability under different loads and cluster sizes.

\subsection{Experiment Setup}

\begin{table}[ht]
\centering
    \begin{tabular}[t]{rp{0.65\linewidth}}
        \toprule
        Resource&Details\\
        \midrule
        OS&Ubuntu 18.04.3\\
        CPU&Quadcore Intel Xeon CPU E3-1230 V2 3.30GHz\\
        Memory&16 GB RAM\\
        Storage&3TB RAID0 (3x1TB disks, linux software RAID)\\
        Network&1 GBit Ethernet NIC\\
        Software&Java v1.11, Flink v1.12, Kafka v2.6, ksqlDB v6.1, RabbitMQ v3.8, ZooKeeper v3.6, Docker v19.3, Kubernetes v1.18, HDFS v2.8,, Prometheus v2.25
        \\
        \bottomrule
    \end{tabular}
\caption{Cluster Specifications}
\label{clusterspecs}
\end{table}%

Our experimental setup consisted of a 50-node co-located Kubernetes and HDFS cluster. 
Node specifications and software versions are summarized in Table \ref{clusterspecs}. A single switch connected all nodes. 
Each experiment consisted of a Kubernetes namespace containing: an Apache Kafka cluster of size 3 with all topics configured to have 8 partitions and a replication factor of 3; an Apache Zookeeper cluster of size 3; a single ksqlDB server; a single RabbitMQ server; an Apache Flink native session cluster\footnote{ https://ci.apache.org/projects/flink/flink-docs-stable/deployment/ resource-providers/native\_kubernetes.html, Accessed: Jul 2021}; and a single Prometheus\footnote{ https://prometheus.io, Accessed: Jul 2021} time series database was used for the gathering of metrics. 
We provide a Github repository\footnote{https://github.com/dos\-group/tdis21\_dsp\_platform} containing the helm chart for our cluster setup as well as the Flink enrichment and analytics jobs.
Experiments were conducted 5 times with the median selected for our results.

\begin{figure*}
    \centering
    \begin{subfigure}[t]{0.48\columnwidth}
        \includegraphics[width=\textwidth]{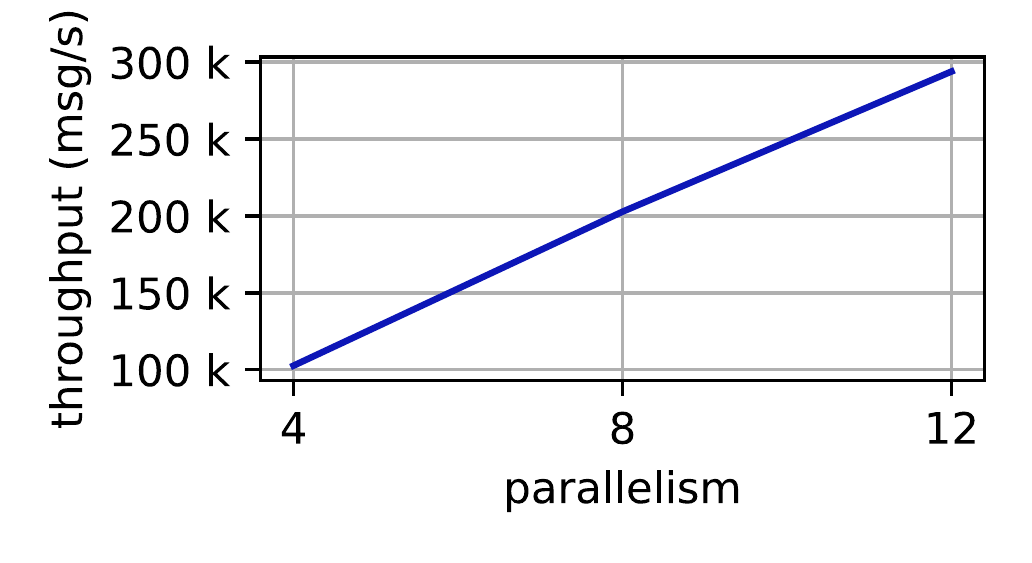}
        \caption{Scalability: Enrichment Job}
        \label{fig:ETL-avail-results}
    \end{subfigure}
    \hfill
    \begin{subfigure}[t]{0.48\columnwidth}
        \includegraphics[width=\textwidth]{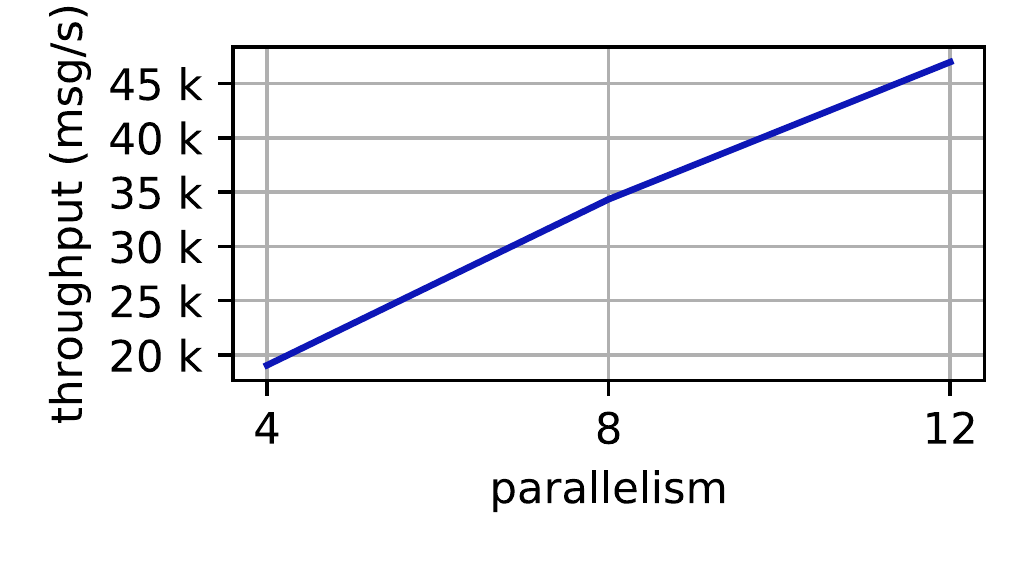}
        \caption{Scalability: Analytics Job}
        \label{fig:CEP-avail-results}
    \end{subfigure}
    \hfill
    \centering
    \begin{subfigure}[t]{0.48\columnwidth}
        \includegraphics[width=\textwidth]{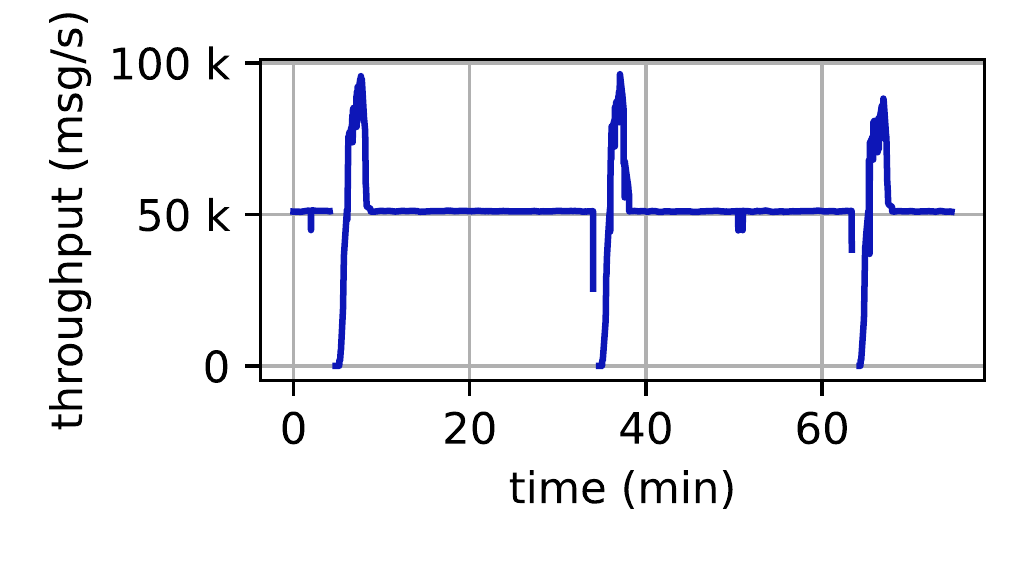}
        \caption{Reliability: Enrichment Job}
        \label{fig:ETL-rel-results}
    \end{subfigure}
    \hfill
    \begin{subfigure}[t]{0.48\columnwidth}
        \includegraphics[width=\textwidth]{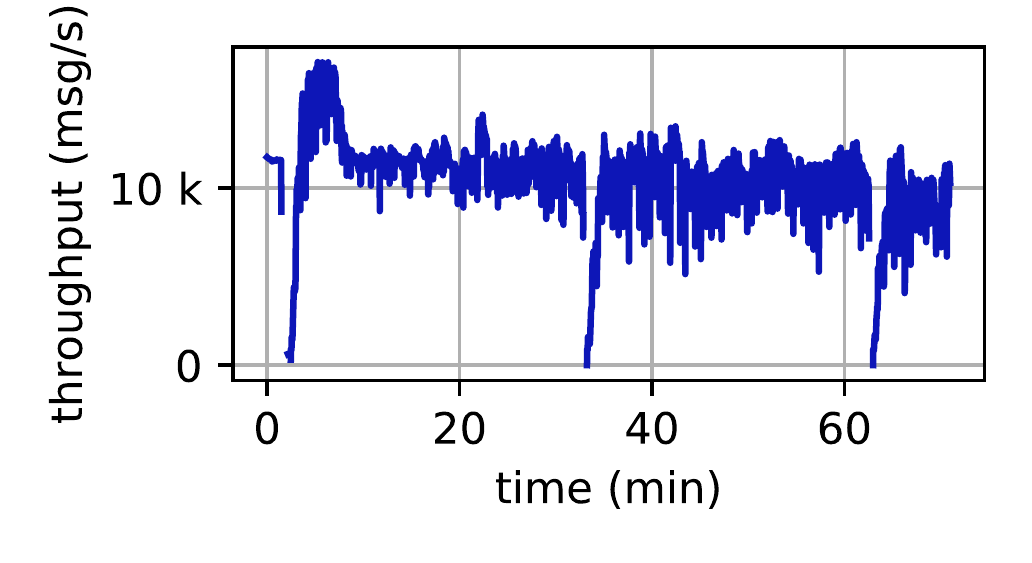}
        \caption{Reliability: Analytics Job}
        \label{fig:CEP-rel-results}
    \end{subfigure}
    \hfill
    \caption{Results of scalability and reliability experiments.}\label{fig:experimental-results}
\end{figure*}

\subsection{Scalability Experiment}

To test the scalability of the IoT data processing platform, we measured the maximum throughput for a data enrichment and an analytics job at varying cluster sizes.
Because stream processing systems will process as many messages as possible using the resources available, the maximum processing capacity can be deduced by ensuring the messaging platform contains more messages than can be processed at a given time.
Utilizing this fact, we generated enough messages to Kafka to sustain the maximum processing capacity for at least 20 minutes at each cluster size.
We then recorded the mean throughput for the different cluster sizes across multiple runs.

From the results of the scalability experiments shown in Fig.~\ref{fig:ETL-avail-results} and Fig.~\ref{fig:CEP-avail-results}, we see that each job is able to process a different number of maximum messages at each cluster size.
However, increasing the cluster size for both jobs resulted in a near-linear increase in processing capacity.
We can furthermore deduce the amount of resources needed to provide dependable service.
For example, with a cluster size of four, the data enrichment job can dependably process 50,000 messages per second, allowing for additional processing capacity in the event of failure.

\subsection{Reliability Experiment}

To test the reliability of the platform, we ran experiments where failures were injected into both the data enrichment and analytics jobs.
Chaos mesh\footnote{https://chaos-mesh.org/, Accessed Jul 2021} was used to inject 3 failures every 30 minutes which caused a random task manager container to fail and the jobs to restart.
We then measured the mean time to recover.
Flink was configured to use a checkpoint interval of 10 seconds, a timeout interval of 50 seconds, and to use \textit{exactly-once} fault tolerance guarantees.
Workloads which oscillated close to 50K and 10K messages per second were used for the enrichment and analytics jobs respectively. 
A parallelism of 4 was used for both jobs resulting in an average utilization rate of 50\% considering maximum processing capacities established in the scalability experiment.

Fig.~\ref{fig:ETL-rel-results} and Fig.~\ref{fig:CEP-rel-results} shows the input throughput over time for these experiments. 
Time taken to restart the job after failures were detected averaged 10 seconds across all jobs.
The average time taken to process the backlog of accumulated events while the system was unavailable, i.e. the consumer lag, was 194 and 243 seconds for the enrichment and analytics jobs respectively.
In all cases the jobs were able to recover and catch up to processing events at the latest timestamp.

\subsection{IoT Analytics Experiment}
For an evaluation of the alerts emitted by the proximity-based CEP job, we created a simulation with six geographical clusters (cells) of sensors, each containing between one and fifty sensors.
We then simulated their measurements with values fluctuating around some mean per cell (with noise) and injected anomalies of magnitudes the job should emit alerts for for one, multiple, or all sensors in a cell.

\begin{table}[ht]
\centering
    \begin{tabular}[t]{rll}
        \toprule
        &Anomalous readings&Normal readings\\
        \midrule
        Associated with alert&34&1689.8\\
        No associated alert&0&27.2
        \\
        \bottomrule
    \end{tabular}
\caption{Results of the location-aware CEP analysis}
\label{tab:cep-qualitative}
\end{table}%

During five runs of our evaluation parcours, we produced three separate anomalous situations in different cells with a total of 12 sensors ever reporting values deviating from the mean.
Table~\ref{tab:cep-qualitative} shows that our system associated alarms to all anomalous readings as well as to $27.2$ normal readings ($1.6\%$ false positive rate).
The compromise between false positives and false negatives can be readjusted by choosing different parameters for the allowable deviation from the mean.
\section{Related Work}

When looking at the current state of research one can find several papers adapting a similar architecture, or parts of it, for solving problems comparable with the ones introduced here.
The paper of Sahal et al.~\cite{sahal_big_2020} examines existing big data technologies for predictive maintenance use cases in the industry 4.0.
Ye at al.~\cite{ye_hydrologic_2020} utilizes the scalability of Flink and sliding window partitioning to detect anomalies in hydrologic time series. A different application is presented in Difallah et al.~\cite{difallah_scalable_2013} where Apache Storm is used to detect anomalies in a water distribution network. More work on the here proposed platform can be found in an earlier work of our research group in Lorenz et al.~\cite{lorenz2020scalable}
Nasiri et al.~\cite{nasiri_evaluation_2019} and Hesse and Lorenz~\cite{hesse_conceptual_2015} study capabilities and architecture of data stream processors. On the topic of message brokers, Moskvicheva and Dolgachev~\cite{moskvicheva_industry_2020} compare RabbitMQ and Apache Kafka as two popular open source options.
Other platform proposals can be found in Huru et al.~\cite{huru_bigclue_2018} and Malek et al.~\cite{malek_use_2017} who developed general purpose platforms for distributed, scalable, and fault tolerant IoT data stream processing. No data about scalability or reliability of these approaches were given though. The paper of Zeuch et al.~\cite{zeuch2020nebulastream} and its fog computing approach is focused on future special cases where centralised processing might not be a viable solution.
Our previous work in the area of dependable stream processing includes optimizing the fault tolerance mechanisms of DSP systems to ensure compliance with user-defined Quality of Service constraints for both performance and availability~\cite{DBLP:conf/bigdataconf/GeldenhuysTGLK19,DBLP:conf/bigdataconf/GeldenhuysTK20}.
Additionally, our work also explores the area of dependable data analytics platforms specific to water infrastructure monitoring~\cite{DBLP:conf/bigdataconf/LorenzGSJLSBT20,HLT21}.

\section{Conclusion}

In this paper we presented a dependable IoT data processing platform used for the monitoring and control of urban infrastructures.
The goal of such a platform is to provide near real-time monitoring of the current state of the infrastructure as well as analytics for geo-distributed use cases such as predictive maintenance. 
The platform is intended to operate in the cloud or locally in a fog-based environment with the ability to adapt to any changes in the workload over time. 
We aimed to present an architecture which is not overly complex and composed of mature open-source software solutions.
In the future we wish to perform further testing of this platform and jobs at greater scales using real world data streams.

\section*{Acknowledgment}

This work has been supported through grants by the German Ministry for Education and Research (BMBF) as WaterGridSense 4.0 (funding mark 02WIK1475D) as well as OPTIMA which is co-financed by the European Regional Development Fund.
We also thank Felix Lorenz for his contributions to this research and all the partners involved in the two funded projects acknowledged here.

\bibliographystyle{IEEEtran}
\balance
\bibliography{paper}

\end{document}